\input epsf
\newcommand{\be}{\begin{equation}}
\newcommand{\bea}{\begin{eqnarray}}
\newcommand{\eea}{\end{eqnarray}}
\newcommand{\ee}{\end{equation}}
\newcommand{\eqn}[1]{\label{#1}}
\newcommand{\eq}[1]{Eq.~(\ref{#1})}

\newcommand{\eqs}[1]{Eqs.\ (\ref{#1})}

\newcommand{\cA}{\mathcal{A}}
\newcommand{\cB}{\mathcal{B}}

\newcommand{\C}{\mathcal{C}}

\newcommand{\cH}{\mathcal{H}}
\newcommand{\xtot}{X^{\mathrm{tot}}}
\newcommand{\ytot}{Y^{\mathrm{tot}}}
\newcommand{\ztot}{Z^{\mathrm{tot}}}
\newcommand{\sigmatot}{\vec{\sigma}^\mathrm{tot}  }

\newcommand{\ad}{+}
\newcommand{\udec}{U_{\mathrm{dec}}}
\newcommand{\tr}{\mathrm{tr}}
\newcommand{\rhotilde}{\tilde{\rho}}

\documentclass[a4paper,pra,twocolumn,nofootinbib,superscriptaddress]{revtex4}

\usepackage{amssymb}
\usepackage{amsmath}
\usepackage{amsthm}
\usepackage{hyperref}

\theoremstyle{plain}
\newtheorem*{proposition}{Proposition}

\begin{document}
\bibliographystyle{apsrev}

\setlength{\arraycolsep}{0.05cm}  %0.05
\title{Optimal state encoding for quantum walks and quantum communication over spin
systems}

\author{Henry L.~Haselgrove}
\email{HLH@physics.uq.edu.au}
 \affiliation{School of Physical
Sciences, University of Queensland, Brisbane 4072, Australia}
\affiliation{Information Sciences Laboratory, Defence Science and
Technology Organisation, Edinburgh 5111 Australia}
\date{\today}
 \begin{abstract}
%\abstract{
 Recent work has shown that a simple chain of interacting spins
can be used as a medium for high-fidelity quantum communication.
 We describe a scheme for quantum communication using a spin system that conserves
 $z$-spin, but otherwise is arbitrary. The sender and receiver are assumed to directly control several
spins each, with the sender encoding the message state onto the
larger state-space of her control spins. We show how to find the
encoding that maximises the fidelity of communication, using a
simple method based on the singular-value decomposition.  Also, we
show that this solution can be used to increase communication
fidelity in a rather different circumstance: where no encoding of
initial states is used, but where the sender and receiver control
exactly two spins each and vary the interactions on those spins
over time. The methods presented are computationally efficient,
and numerical examples are given for systems having up to 300
spins.

%}
\end{abstract}

\maketitle

 \section{Introduction}

Quantum communication, the transfer of a quantum state from one
place or object to another, is an important task in quantum
information science\cite{DiVincenzo00a}. The problem of
communicating quantum information is profoundly different to the
classical case\cite{Preskill98c,Nielsen00a}. For example, quantum
communication could not possibly be achieved by just measuring an
unknown state in one place, and reconstructing in another. Rather,
an entire system of source, target, and medium must evolve in a
way that maintains quantum coherence.

In this paper we consider an idealised system of interacting
spin-1/2 objects, isolated from the environment. The aim is to use
the system's natural evolution to communicate a qubit state from
one part of the system to another. The motivation is that such a
system could be used as a simple ``quantum wire'' in future
quantum information-processing devices. The most obvious
configuration to choose is a simple one-dimensional open-ended
chain, with interactions between nearest-neighbour spins, in which
case we want the chain's evolution to transfer a qubit state from
one end to the other. The methods in this paper apply to this
simple type of chain, and also to spin networks of arbitrary
graph.

A number of interesting proposals exist for quantum communication
through spin chains. In \cite{Bose02a}, the 1D Heisenberg chain
was considered, with coupling strengths constant over the length
of the chain and with time. The idea was to initialise all spins
in the ``down'' state, except the first spin, which was given the
state of the qubit to be sent. After the system was allowed to
evolve, the spin at the far end of the chain would then contain
the sent state, to some level of fidelity. Simulations were
carried out for a range of chain lengths, and it was shown that
the fidelity was high only for very small chains.

In \cite{Christandl03a}, a 1D spin chain with $XY$ couplings was
considered. Here, the coupling strengths were constant over time,
but were made to vary over the length of the chain in a specific
way. Like \cite{Bose02a}, the first spin was initialised in the
state to be sent, with all other spins initialised to ``down''. It
was shown that this scheme allows a {\em perfect} state transfer
to the far spin site, for any length of chain.

In \cite{Osborne03b}, a scheme was presented for high-fidelity
quantum communication over a {\em ring} of spins with
nearest-neighbour Heisenberg couplings, using coupling strengths
constant over the length of the ring and over time. The sender and
receiver are located diametrically opposite to one another. The
authors showed that excitations travel around the ring in a way
that can be described using a concept from classical wave theory,
the dispersion relation. Using this insight, they constructed a
scheme where the sender, who controls several adjacent spins,
constructs an initial state that is a Gaussian pulse having a
particular group velocity chosen to minimise the broadening of the
pulse over time. Using this state for the encoding of the
$|1\rangle$ basis of the qubit message, and the all-down state as
the encoding of $|0\rangle$, an arbitrary qubit can be sent with
high fidelity over rings of any size, so long as the number of
spins that the sender controls is at least the cube root of the
total number.

Motivated by the results in \cite{Osborne03b}, we pose the
following problem. Say we are given the Hamiltonian for a system
of interacting spins, where the graph of the interactions is not
necessarily a ring structure, but is completely arbitrary. Also,
the strength and type of interaction along each graph edge is
arbitrary (so long as total $z$ spin is conserved). The sender
Alice controls some given subset of the spins, and the receiver
Bob controls some other given subset. How does Alice encode the
qubit to be sent onto the spins she controls, in order to maximise
the fidelity of communication? We know from \cite{Osborne03b} that
the Gaussian pulse provides a near-optimal fidelity for the case
of a Heisenberg ring (and is optimal in the limit of large ring
sizes). What about other shapes of spin network? Can we find a
general solution?

We provide a simple and efficient method for finding the
maximum-fidelity encoding of the $|1\rangle$ message basis state,
for a general $z$-spin-conserving spin system. (We assume that the
encoding for the $|0\rangle$ basis state is fixed to the all-down
state). So, unlike the schemes in \cite{Bose02a},
\cite{Christandl03a}, and \cite{Osborne03b}, which use systems
with interactions that have specific strengths and conform to a
specific graph, our scheme is designed to ``make the most'' of
whatever arbitrary system is given to us. We give a numerical
example of our method, for a system of 300 spins (where Alice and
Bob each control 20 spins), showing a near-perfect average
fidelity.

 We give a second scheme for increasing fidelity, that does not
use encoding of initial states, but relies on Alice and Bob
dynamically controlling the interactions on their control spins.
Here, the number of control spins is fixed at two each for Alice
and Bob. We give a straightforward method for deriving control
functions, that give a fidelity (and communication time) equal to
the values that would result if Alice and Bob had instead each
controlled many more spins (with static interactions) and used the
optimal initial-state encoding scheme. This method has the
combined benefits of being applicable to arbitrary
$z$-spin-conserving spin-chains, yet having a fixed two-spin
``interface'' with Alice and Bob. We give numerical examples, and
plot the derived control functions, for a 104-spin and a 29-spin
system,  showing a near-perfect fidelity in each case.

In the remainder of this introductory section, we briefly describe
the assumptions behind our schemes, and define our notation.
Sec.~\ref{firstsec} describes our method of deriving the optimal
message encoding. Sec.~\ref{secondsec} describes our scheme for
increasing fidelity via dynamic control. Concluding remarks are
made in Sec.~\ref{conclusion}.

\subsection{Assumptions and notation}

The solution presented in this paper relies on two main
assumptions, which we now list. Firstly, the system Hamiltonian
must commute with $\ztot$, which we define to be the $z$-component
of the total spin operator
\begin{equation}
\sigmatot \equiv \left(\xtot,\ytot,\ztot\right) \equiv \sum_j
\vec{\sigma}_j,
\end{equation}
where $\vec{\sigma}_j$ is the vector of Pauli operators
$(\sigma^x,\sigma^y,\sigma^z)$ acting on the $j$-th spin. The
Pauli operators in the basis ``down'' $|\downarrow\rangle$ and
``up'' $|\uparrow\rangle$ are
\begin{equation} \sigma^x=\left[\begin{array}{rr} 0&1\\1&\phantom{-}0
\end{array} \right];
\quad
 \sigma^y=\left[\begin{array}{rr} 0&-i\\i&0 \end{array}\right]
 ;\quad
 \sigma^z=\left[\begin{array}{rr} 1&0\\0&-1 \end{array}
\right].
\end{equation}
Secondly, the spin system must be initialised to the all-down
state $|\downarrow\rangle \otimes \dots \otimes
|\downarrow\rangle$, before the communication is carried out. Note
that the schemes in \cite{Bose02a},\cite{Christandl03a} and
\cite{Osborne03b} also make use of these two assumptions. It could
be argued that the first condition is reasonable because it
follows from rotational invariance. Of course, any external
magnetic field will destroy this invariance, and in particular any
magnetic field which is not in the $z$ direction will mean that
the $z$-component of total spin is no longer conserved. The
Heisenberg and $XY$ interactions are examples of interactions that
conserve $\ztot$. The second constraint might be rather difficult
to achieve in practice. One possibility would be to apply a strong
polarising magnetic field in the $z$ direction, over the entire
system, and let the system relax to its ground state.

 In the remainder of the paper, in place of the notation $|\downarrow\rangle$ and
$|\uparrow\rangle$ for the eigenstates of $\sigma^z$, we will use
the equivalent but more convenient notation $|0\rangle$ and
$|1\rangle$. A {\em computational basis state} of the system is
defined to be one where each spin is in either a $|0\rangle$ state
or a $|1\rangle$ state. Note that the computational basis states
are all eigenstates of $\ztot$, and the eigenvalue has one of
$N+1$ possible values, given by the number of $|0\rangle$s minus
the number of $|1\rangle$s. So in a system of $N$ spin-1/2
objects, we can break the state space into $N+1$ subspaces of
different well-defined $z$-component of total spin. We use
$\cH^{(n)}$, $n=0,\dots,N$, to denote these subspaces. $\cH^{(n)}$
is the eigenspace of $\ztot$ that is spanned by the $N \choose n$
computational basis states that have $n$ qubits in the $|1\rangle$
state and the rest in the $|0\rangle$ state.

Since the system Hamiltonian $H$ commutes with $\ztot$, a state in
$\cH^{(n)}$ will remain in $\cH^{(n)}$ under the evolution of $H$.
$\cH^{(0)}$ is one-dimensional; it is spanned by the all-zero
state $|0\rangle \otimes $\dots$\otimes|0\rangle$. So this state
is a stationary state of $H$.

\section{The optimal encoding scheme} \label{firstsec}

Say that Alice wishes to send the qubit state $\alpha|0\rangle +
\beta |1\rangle$. In our scheme, she does so by preparing the
state $\alpha |\mathbf{0}\rangle_A + \beta |1_{ENC}\rangle_A$,
where $|\mathbf{0}\rangle_A$ is the all-zero state on her spins,
and $|1_{ENC}\rangle_A$ is some state orthogonal to
$|\mathbf{0}\rangle_A$ (the ``{\em ENC}'' stands for ``encoded'').
(Note that Alice doesn't necessarily know $\alpha$ and $\beta$.
She would presumably prepare the state by some unitary operation
acting on her spins and some external spin containing the state
$\alpha|0\rangle+\beta|1\rangle$.) We assume that the entire spin
chain is initialised to the all-zero state, so immediately after
Alice prepares the abovementioned state on her spins, the state of
the whole system is
\begin{equation}
|\Psi(0)\rangle \equiv (\alpha |\mathbf{0}\rangle_A + \beta
|1_{ENC}\rangle_A) \otimes |\mathbf{0}\rangle_{\bar{A}},
\eqn{init}
\end{equation}
where $\bar{A}$ refers to all spins that Alice does not control.
The whole spin system is allowed to evolve for a time $T$, giving
the state $|\Psi(T)\rangle=e^{-iHT}|\Psi(0)\rangle$. Using the
fact that $|0\rangle\otimes\dots\otimes|0\rangle$ is a stationary
state, $|\Psi(T)\rangle$ can be written (up to some global phase)
as
\begin{eqnarray}
|\Psi(T)\rangle &=& \beta
\sqrt{1-\C_B(T)}|\eta(T)\rangle + \nonumber\\
&&
\hspace{0.5cm}|\mathbf{0}\rangle_{\bar{B}}(\alpha|\mathbf{0}\rangle_B+\beta\sqrt{\C_B(T)}|\gamma(T)\rangle_B)
\eqn{fin} ,
\end{eqnarray}
for some nonnegative $\C_B(T)$, some normalised
$|\gamma(T)\rangle_{B}$ orthogonal to $|\mathbf{0}\rangle_B$, and
for some normalised $|\eta(T)\rangle$ that is orthogonal to all
states of the form $|0\rangle_{\bar{B}}\otimes|v\rangle_B$.

We now show that $\C_B(T)$ can be used as a measure of success.
Comparing Eqs.\ (\ref{init}) and (\ref{fin}), we see that
$\C_B(0)=0$. If $\C_B(T)$ reaches 1 for some later $T$, a
perfect-fidelity quantum communication has resulted. This is
because Bob will then have the state
$\alpha|\mathbf{0}\rangle_B+\beta|\gamma\rangle_B$ on the qubits
he controls, which can be ``decoded'' by a unitary operation into
the state $\alpha|0\rangle + \beta |1\rangle$ of a single spin,
since $|\mathbf{0}\rangle_B$ and $|\gamma\rangle_B$ are
orthogonal. If $\C_B(T)$ is less than $1$, the unitary decoding by
Bob will leave him with a qubit state $\rho$ that is generally
different to the message state. That is, the measure of state
fidelity $F\equiv( \alpha|0\rangle + \beta |1\rangle )^\dagger
\rho ( \alpha|0\rangle + \beta |1\rangle )$ between the message
$\alpha|0\rangle +\beta|1\rangle$ and $\rho$, will generally be
less than one whenever $\C_B(T)<1$. However, the value of $F$ is
highly dependent on the message state
--- for example, if $\alpha=1$ then $F=1$ regardless of the value
of $\C_B(T)$.

$\C_B(T)$, on the other hand, is a message-independent measure of
the fidelity of communication. Consider $\bar{F}$, defined to be
the state fidelity $F$ averaged over all message states. For
encodings $|1_{ENC}\rangle$ that belong to the $\cH^{(1)}$
subspace, we have
\begin{equation}
 \bar{F}=\frac{1}{2}+\frac{1}{3}\sqrt{\C_B(T)} +
\frac{1}{6}\C_B(T), \eqn{fbar}
\end{equation}
which is a monotonic function of $\C_B(T)$ \cite{Osborne03b}. So,
in this case maximising the {\em average} state fidelity is
equivalent to maximising $\C_B(T)$. More generally, for
$|1_{ENC}\rangle$ not in $\cH^{(1)}$, the expression in \eq{fbar}
provides a reasonably tight lower bound on $\bar{F}$:
\begin{eqnarray}
\frac{1}{2}+\frac{1}{3}\sqrt{\C_B(T)} + \frac{1}{6}\C_B(T) \leq
\bar{F} &\leq& \frac{1}{2}+\frac{1}{3}\sqrt{\C_B(T)} + \frac{1}{6}
\nonumber\\
&=&\frac{2}{3}+\frac{1}{3}\sqrt{\C_B(T)}.
\end{eqnarray}
The precise value of $\bar{F}$ will then depend on
$|\eta(T)\rangle$ and the full specification of Bob's decoding
unitary.

A further argument for using $\C_B(T)$ as a measure of
communication fidelity comes from considering the system's ability
to transfer {\em quantum entanglement} from Alice to Bob. Suppose
that Alice, instead of sending a message which is a pure quantum
state $\alpha|0\rangle + \beta|1\rangle$, sends a state which is
maximally entangled with some additional spin that Alice
possesses. (The additional spin does not interact when the system
evolves). If the communication is perfect, the result must be that
Bob's decoded message becomes maximally entangled with Alice's
additional spin. So more generally, when the communication is not
perfect, the amount of entanglement generated between Alice and
Bob would be a good measure of communication fidelity. In fact,
the entanglement generated, measured by the {\em concurrence}, is
equal to $\sqrt{\C_B(T)}$ (a proof of this fact is outlined in
Appendix A). This is independent of $|\eta(T)\rangle$, or the full
specification of Bob's decoding unitary, or whether
$|1\rangle_{ENC}$ belongs to $\cH^{(1)}$.

To recap, when the Hamiltonian commutes with $\ztot$ and the state
is initialized to $|\mathbf{0}\rangle$, the problem of achieving a
high communication fidelity can be boiled down to choosing an
appropriate initial encoding $|1_{ENC}\rangle$ for the $|1\rangle$
qubit basis state. We seek a state $|1_{ENC}\rangle_A \otimes
|\mathbf{0}\rangle_{\bar{A}}$ that has the property that it
evolves to (or near to) a state of the form
$|\mathbf{0}\rangle_{\bar{B}} \otimes |\gamma\rangle_B$, or in
other words such that $\C_B(T)\approx 1$ for some $T$. Alice's
choice for the ``encoding'' of the $|0\rangle$ qubit basis state
is fixed to the all-zero state. With perfect fidelity that basis
state will evolve to the all-zero state on Bob's spins. (Note that
in some cases it may be possible to increase fidelity further by
allowing an encoding for $|0\rangle$ other than the all-zero
state. We ignore such a possibility, in order to keep the method
for finding the encoding simple and efficient. The simplification
is used likewise in \cite{Osborne04a}.)

 We now show that the encoding $|1_{ENC}\rangle$ which maximises
 $\C_B(T)$ for a given $T$ can be found by performing the
 singular value decomposition on a modified version of the evolution
 matrix $e^{-iHT}$.

 Let $\cA$ be the vector subspace of states of the form
$|1_{ENC}\rangle_A \otimes |\mathbf{0}\rangle_{\bar{A}}$, such
that $_A\langle\mathbf{0}|1_{ENC}\rangle_A=0$. Similarly, let
$\cB$ be the vector subspace of states of the form
$|\mathbf{0}\rangle_{\bar{B}} \otimes |\gamma\rangle_B$, such that
$_B\langle\mathbf{0}|\gamma\rangle_B=0$. In other words, $\cA$
reflects all the possible encodings that Alice could use for the
$|1\rangle$ qubit basis state (regardless of the fidelity they
would achieve). $\cB$ is the set of states that we would like some
state in $\cA$ to evolve to; a state in $\cA$ that evolves to one
in $\cB$ represents an encoding for $|1\rangle$ that gives
$\C_B(T)=1$ and thus a perfect average fidelity.

 Let $P_\cA$ and
$P_\cB$ be the projectors onto the subspaces $\cA$ and $\cB$. Let
$U(T)\equiv e^{-iHT}$ be the time-evolution operator.
 From \eqs{init} and (\ref{fin}), we can write
\begin{equation}
\C_B(T)= \| \hspace{1mm} P_\cB U(T) |1_{ENC}\rangle_A {\otimes}
|\mathbf{0}\rangle_{\bar{A}} \hspace{1mm} \|^2,
\end{equation}
where $\| \cdot \|$ denotes the $l_2$-norm. This means that for a
particular total communication time $T$, choosing the optimal
initial encoding for the $|1\rangle$ state is a matter of finding
the normalised $|\psi\rangle \in \mathbb{C}^{2^N}$ that maximises
$\| P_\cB U(T) P_\cA |\psi\rangle \|$. The maximum value is given
by the largest singular value of $\tilde{U}(T)\equiv P_\cB U(T)
P_\cA $, and the corresponding optimal $|\psi\rangle$ is the first
right-singular-vector of $\tilde{U}(T)$ \cite{Horn91a}. Recall,
the SVD (singular value decomposition) of $\tilde{U}(T)$ is
\begin{eqnarray}
\tilde{U}(T)&=&V S W^\dagger \\
&=& \left( %
\begin{array}{ccc}
\vec{v}_1 \hspace{0.5mm} & \vec{v}_2  \hspace{0.5mm} & \ldots \\
\downarrow & \downarrow & \ldots \\
{\rule{0em}{1.55em}} & {} & {}
\end{array}
 \right) %
 \left( %
\begin{array}{ccc}
s_1 & {} & {} \\
{} & s_2 & {} \\
{} & {} & \ddots
\end{array}
 \right) %
  \left( %
\begin{array}{ccc}
\vec{w}_1^* & \rightarrow & {\hspace{3mm}} \\
\vec{w}_2^* & \rightarrow & {} \\
\vdots & \vdots & {}
\end{array}
 \right), \nonumber
\end{eqnarray}
where $s_1 \ge s_2 \ge \dots \ge 0$ are the {\em singular values},
the orthonormal $\vec{w}_j$ are the {\em right singular vectors},
and the orthonormal $\vec{v}_j$ are the {\em left singular
vectors} of $\tilde{U}(T)$. Numerical packages such as Matlab have
in-built routines for easily calculating the SVD. So, we have that
$\C_B(T)$ has its maximum value, $s_1^2$, when Alice chooses the
initial state $|1_{ENC}\rangle_A \otimes
|\mathbf{0}\rangle_{\bar{A}}=\vec{w}_1$ to encode $|1\rangle$.

Other parts of the decomposition could be useful as well. Say
Alice wants to transmit two qubits {\em simultaneously} to Bob. If
she uses the all-down state to encode the $|00\rangle$ basis
state, then she should use the encodings $\vec{w}_1$, $\vec{w}_2$,
and $\vec{w}_3$ for the other three basis states $|01\rangle$,
$|10\rangle$, and $|11\rangle$. Then, so long as $s_3\approx 1$,
the two qubits would be simultaneously communicated with high
fidelity.

The vectors $\vec{w}_j$ and the values $s_j$ are also the
eigenvectors and square-root eigenvalues respectively of
$(P_B\tilde{U}(T)P_A)^\dagger P_B\tilde{U}(T)P_A $ = $P_A
\tilde{U}^\dagger(T) P_B \tilde{U}(T) P_A$. Now, $\ztot$ commutes
with $P_A \tilde{U}^\dagger(T) P_B \tilde{U}(T) P_A$ because it
commutes with each of $P_A$, $P_B$, $\tilde{U}(T)$, and
$\tilde{U}^\dagger(T)$ separately. So the $\vec{w}_j$ will all
have well-defined total $Z$ spin (or can be chosen to, wherever
ambiguities exist because of degeneracies in the $s_j$). This is
important when it comes to calculating these solutions
efficiently. Instead of performing the full $2^N$ by $2^N$ matrix
exponential and SVD, the calculation can be done separately for
each of the smaller subspaces $\cH^{(n)}$, starting each
calculation with the $N \choose n$-by-$N \choose n$ part of the
Hamiltonian that acts on the $\cH^{(n)}$ subspace.

 Alice can't create a state with more than $|A|$ qubits
in the ``one'' state, where $|A|$ is the number of spins she
controls. So, in fact the calculation only needs to be done over
the $\cH^{(1)}$, \dots{ }, $\cH^{(|A|)}$ subspaces (in other
words, the singular values corresponding to states in other
subspaces will always be zero).

In practice we have found that the optimal solution $\vec{w}_1$
often belongs to the $\cH^{(1)}$ subspace. (In particular, we
calculated the optimal solution for a range of different values of
$T$ for various 8 and 9-spin systems, and found that only for a
very small minority of the values of $T$, for each system, was the
solution {\em not} in the $\cH^{(1)}$ subspace ).
 A rudimentary argument
for this can be made as follows. Looking for solutions in
$\cH^{(n)}$ means optimising over Alice's ${|A|} \choose n$
degrees of freedom (of the space $\cA \cap \cH^{(n)}$), in order
to make the final state land in or near a ${|B|} \choose
n$-dimensional target space $\cB \cap \cH^{(n)}$. This must be
achieved despite the fact that the Hamiltonian is ``trying'' to
move the state through a much larger $N \choose n$-dimensional
space $\cH^{(n)}$. Over the various values of $n=1,\dots,|A|$, the
dimensionality of $\cA \cap \cH^{(n)}$ and $\cB \cap \cH^{(n)} $
as a fraction of the dimensionality of $\cH^{(n)}$ is largest when
$n=1$. In other words, when $n=1$, the size of the target space,
and amount of control available of the initial state, is largest
as a fraction of the dimensionality of the entire subspace
$\cH^{(n)}$.

So, in general we can restrict all the calculations to the
$N$-dimensional subspace $\cH^{(1)}$, and there will still be a
good chance that we will arrive exactly at the globally-optimal
encoding $\vec{w}_1$. Ignoring solutions in the other subspaces
will increase the efficiency of calculation considerably,
especially for large chains.

The evolution of a state in the $\cH^{(1)}$ subspace can also be
interpreted as a {\em continuous quantum walk} of a particle over
a graph. ( For an introduction to quantum walks, see for example
\cite{Kempe03b} and references therein).  The graph is simply the
graph of interactions between spins in the Hamiltonian $H$, and
the state $|\mathbf{0}\rangle_{\bar{j}}\otimes|1\rangle_j$
corresponds to the particle being at vertex $j$ of that graph. So
our methods for increasing communication fidelity are,
equivalently, methods for guiding a quantum walk from one part of
a graph to another. We point out this connection because of the
significant interest currently in using quantum walks for solving
computational problems (see for example
\cite{Childs03c,Ambainis03a,Osborne04a} and references therein).

%\subsection{Example: open-ended Heisenberg chain}

To demonstrate the use of the SVD optimal-encoding technique, we
now consider a numerical example. Imagine that Alice and Bob are
joined by a $300$-site open-ended chain with nearest-neighbour
couplings given by the antiferromagnetic Heisenberg interaction,
with coupling strengths all equal to 1. That is,
\begin{equation}
H=\sum_{j=1}^{299} \vec{\sigma}_j\cdot\vec{\sigma}_{j+1}.
\eqn{hexamp}
\end{equation}
Assume that Alice and Bob control the first and last 20 spins
respectively.

In light of the earlier discussion, we restrict our optimisation
to the $\cH^{(1)}$ subspace, and thus ignore all singular vectors
in other subspaces. A Matlab program is used to carry out the
following calculations. First, the 300 by 300 matrix $H^{(1)}$,
defined to be the part of $H$ that acts on $\cH^{(1)}$, is
constructed. Then, the SVD of
\begin{equation}
\tilde{U}^{(1)}(T)=P_{\cB \cap \cH^{(1)}} e^{-i H^{(1)} T} P_{\cA
\cap \cH^{(1)}}
\end{equation}
 is calculated for a range of values of $T$. The optimal value for
communication time is not known beforehand, so this repetition of
the calculation for different values of $T$ is needed in order to
find a reasonable tradeoff between communication time and
fidelity.

\begin{figure}%[!tbp]
\begin{center}
\epsfxsize=8cm \epsfbox{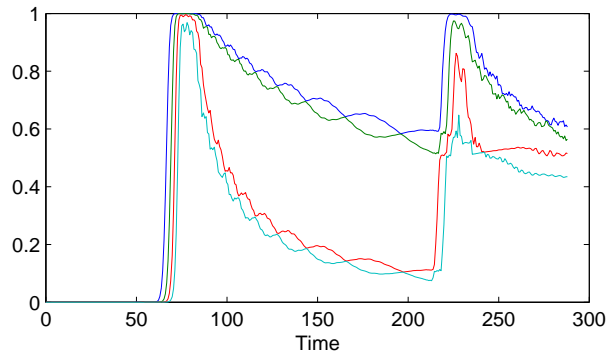} \caption{The largest four
singular values of $\tilde{U}^{(1)}(T)$, for a range of
communications times $T$. Open-ended Heisenberg chain, $N$=300,
and Alice and Bob each control 20 sites.}\label{fig:a}
\end{center}
\end{figure}

The four largest singular values, $s_1$, \dots, $s_4$, of
$\tilde{U}^{(1)}(T)$ are plotted in Figure \ref{fig:a}. Over the
range of $T$ shown, $s_1(T)$ has its maximum of $0.99999$ at $T =
75.75$. So, this system can transmit a qubit with near-perfect
fidelity, over a time interval of $75.75$. The graph shows that
$s_2(75.75)$ and $s_3(75.75)$ are also very close to 1, so in fact
{\em two qubits} could be transmitted simultaneously with high
fidelity in this example, using the encodings $|\bf{0}\rangle$,
$\vec{w}_1(75.75)$, $\vec{w}_2(75.75)$ and $\vec{w}_3(75.75)$ for
the two-qubit basis states.

Let's look at the actual optimally encoded states that are
generated in this example. We visualise a state in $\cH^{(1)}$ by
plotting the square magnitude of the coefficients $\psi_j$, where
$\psi_j$ is the coefficient of the basis state that has the $j$-th
spin in the $|1\rangle$ state:
\begin{equation}
\cH^{(1)} \ni |\psi\rangle = \sum_{j=1,\dots,N} \psi_j
\hspace{1mm} |1\rangle_j \otimes |\mathbf{0}\rangle_{\bar{j}}.
\eqn{expansion}
\end{equation}

\begin{figure}%[!tbhp]
\begin{center}
\epsfxsize=8cm \epsfbox{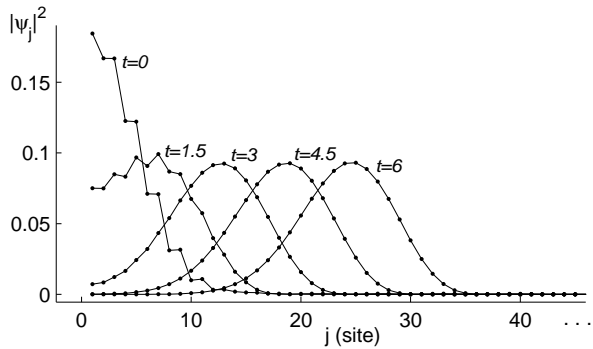} \caption{The optimal
encoded state $\vec{w}_1(75.75)$, evolved for a sequence of times
$t$.}\label{fig:b}
\end{center}
\end{figure}

In Figure \ref{fig:b} we show the evolution of the state
$\vec{w}_1(75.75)$. That is, we set
$|\psi(0)\rangle=\vec{w}_1(75.75)$, and
$|\psi(t)\rangle=e^{-iH^{(1)}t} |\psi(0)\rangle$, and plot the
magnitudes $|\psi_j|^2$ for a sequence of equally-spaced times
$t$. As is necessarily the case, the $t=0$ state has non-zero
coefficients only on Alice's spins, $j=1\dots20$. The state
deforms itself into a Gaussian shape quite quickly. This is
interesting in comparison with the results in \cite{Osborne03b}.
Whilst Gaussian initial states were shown to optimise fidelity on
a Heisenberg ring, the best initial states for an open-ended
Heisenberg chain are ones that deform into a Gaussians. From the
total communication time in this example, the group velocity of
the pulse is roughly $3.95$ (defining the distance between
neighbouring spins to be 1). Thus, in the open-ended Heisenberg
chain we have found the same phenomena that appeared in the
Heisenberg ring in \cite{Osborne03b}, notably that the system has
a preferred group velocity that gives a minimum dispersion and
thus maximum fidelity. This explains the fact that in Figure
\ref{fig:a} the singular values drop for $T$ greater than $75.75$,
and rise again to a near-maximum at $T\approx 225 \approx 3 \times
75.75$: the high-fidelity communication for $T\approx 225$ is also
operating at the preferred group velocity, but the wave packet is
traversing the chain three times, after bouncing from each end.

\begin{figure}%[!tbhp]
\begin{center}
\epsfxsize=8cm \epsfbox{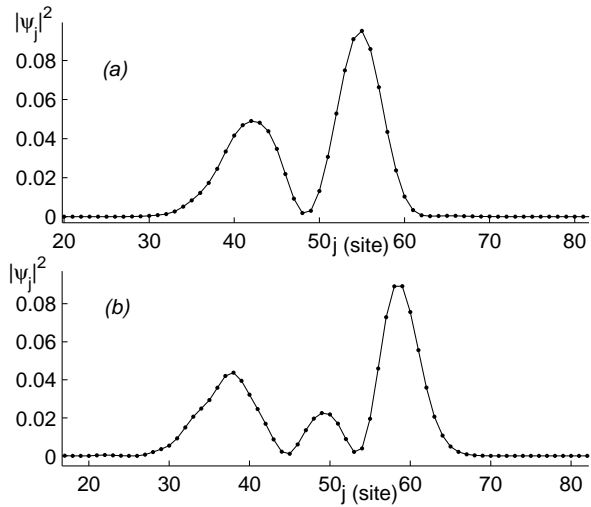} \caption{The states $(a)$
$\vec{w}_2(75.75)$ and $(b)$ $\vec{w}_3(75.75)$ evolved for
time=12. }\label{fig:c}
\end{center}
\end{figure}

Curiously, the lower solutions $\vec{w}_2(75.75)$ and
$\vec{w}_3(75.75)$ seem to evolve into a sum of two and three
Gaussians respectively (see Figure \ref{fig:c}). Animations of
these evolutions are available online \cite{Haselgrove04z}.

%
% Modulated control
%
\section{Dynamic control}  \label{secondsec}

The scheme we presented in Sec.~{\ref{firstsec}} utilises the
evolution of a system having completely static interactions. The
control that Alice and Bob have over the chain is only for an
instant at the beginning and end of the procedure, and so their
only degrees of freedom for increasing the fidelity lie in the
encoding they use. For very long chains, the number of control
sites needed to give a high fidelity might become impractically
large, as suggested by the results in \cite{Osborne03b}. In this
section, we consider the advantage that can be gained by allowing
Alice and Bob to control their spins throughout the procedure, by
modulating the strength of interactions on those spins. The
advantages of the scheme are that the number of control spins are
fixed at four, and that suitable functions for Alice and Bob to
use to vary the interaction strengths are easily derived from a
simple extension of the SVD approach already described.

This type of control scheme is an example of a fundamental problem
in quantum information processing, that of determining how to use
the limited physical control that one has of a quantum system, in
such a way as to achieve the dynamics that are required. For the
task at hand, our method provides a practical and efficient way of
finding an appropriate dynamical control.

\begin{figure}%[!tbp]
\begin{center}
\epsfxsize=8cm \epsfbox{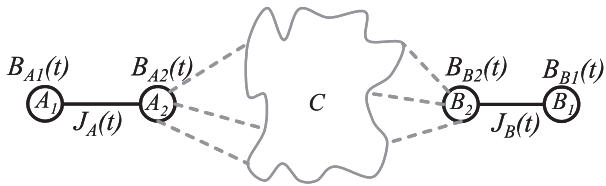}
 \caption{
The general setup, whereby Alice and Bob modulate a total of six
parameters of the Hamiltonian in order to increase communication
fidelity.
 }\label{fig:d}
\end{center}
\end{figure}

A general schematic for the system is shown in Figure \ref{fig:d}.
Alice is now in control of just two spins, labeled $A_1$ and
$A_2$, and, likewise, Bob controls two spins $B_1$ and $B_2$. All
the other spins in the system are collectively denoted $C$. The
graph of the interactions connecting Alice and Bob's spins can be
arbitrary, except that $A_1$ must directly couple only to $A_2$,
and $B_1$ must directly couple only to $B_2$. Like the previous
scheme, we require that the system Hamiltonian $H$ commutes with
$\ztot$, and that all spins are initialised to $|0\rangle$ before
the procedure starts. So again the problem is that of finding a
way of sending the $|1\rangle$ basis state with high fidelity.

The protocol works as follows. At time $t=0$, Alice transfers the
qubit state she wishes to send onto the spin $A_1$. Then she
varies the coupling strength between $A_1$ and $A_2$ according to
some function which we denote $J_A(t)$, and varies the $z$
magnetic field on $A_1$ and $A_2$ according to functions
$B_{A1}(t)$ and $B_{A2}(t)$. At the same time, Bob varies his
coupling strength and $z$ magnetic fields according to the
functions $J_B(t)$, $B_{B1}(t)$ and $B_{B2}(t)$. This process is
continued over some time interval $0<t<T$. At $t=T$ Bob's spin
$B_1$ will contain the sent qubit state, to a level of fidelity
that depends on the six control functions.

\begin{figure}%[!tbhp]
\begin{center}
\epsfxsize=8cm \epsfbox{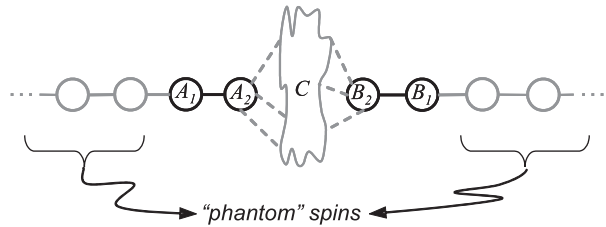}
 \caption{
This is a modified version of Figure (\ref{fig:d}). Calculating
the optimal initial encoded state on this system will help Alice
and Bob derive suitable control functions for their actual
physical setup in Figure (\ref{fig:d}).
 }\label{fig:e}
\end{center}
\end{figure}

How do we choose the control functions in a way that gives us a
high average communication fidelity?
%Surprisingly, the SVD method
%described earlier can be applied here.
The trick is to imagine a
modified version of the system, where a number of ``phantom''
spins have been added to both Alice and Bob's set of control
spins, but all couplings are now fixed (see Figure \ref{fig:e}).

The SVD method is applied to this modified system, to find the
optimal initial state on Alice's extended set of control spins.
The evolution of the encoded state through the modified system is
then simulated on a (classical) computer, and the results of the
simulation are used to derive appropriate control functions for
the actual physical system, using a method which we describe
below. Since, over the bulk of the physical system, the initial
state and interactions are identical to those at the corresponding
regions of the modified system, the problem reduces to finding
control functions which make the state on $A_2$ and $B_2$ in the
physical system evolve in the same way as those corresponding
spins in the modified chain. When that is achieved, the state on
the bulk of the physical chain will evolve in the same way as in
the modified chain, which means that we can communicate a qubit
with the same fidelity as for the optimally encoded state in the
modified system.

For the sake of clarity, we describe the method in detail for a
less general configuration, the 1D $XY$ chain. The derivation is
simpler in this case because, as we shall see, the magnetic
control is not needed. The system Hamiltonian is given by
\begin{eqnarray}
H(t)&=&\sum_{j=2}^{N-2} J_j \left[ \sigma^x_j \sigma^x_{j+1} +
\sigma^y_j \sigma^y_{j+1}\right] \nonumber \\
&& \hspace{0.5cm} + J_A(t)\left(
\sigma^x_1\sigma^x_2+\sigma^y_1\sigma^y_2\right)  \nonumber \\
 && \hspace{0.5cm} + J_B(t) \left(
\sigma^x_{N-1}\sigma^x_N+\sigma^y_{N-1}\sigma^y_N \right) .
\eqn{orig}
\end{eqnarray}
That is, there are some arbitrary fixed $J_j$ that specify the
strengths of the $XY$ couplings over the bulk of the chain. The
strengths at the first and last links can be varied over time by
Alice and Bob.

We write down a Hamiltonian $\tilde{H}$ of a modified system,
where we have extended the length of the chain in both directions
by adding $N_P$ phantom spins to both Alice and Bob's sides. The
new coupling strengths are chosen to be $1$, and the two strengths
that were time-varying in the original system are now also fixed
at one. So that we can use the same numbering system for the spins
as in \eq{orig}, we let the indices of the spins range into the
negative in the modified system, running from $1-N_P$ to $N+N_P$.
The modified Hamiltonian is written simply as
\begin{equation}
\tilde{H}=\sum_{j=1-{N_P}}^{N+N_P-1} J_j \left[ \sigma^x_j
\sigma^x_{j+1} + \sigma^y_j \sigma^y_{j+1}\right],
\end{equation}
where we have extended the definition of $J_j$ so that it equals
$1$ for $j=(1-N_P),\dots, 1$ and for $j=(N-1),\dots,(N+N_P-1)$.

Next, using the SVD method, we find the best encoded initial state
on the set of spins from index $(1-N_P)$ to $1$, while assuming
that the target set of spins ranges from $N$ to $(N+N_P-1)$. That
is, we are imagining a ``modified Alice'' that controls the first
$(N_P+1)$ spins and a ``modified Bob'' that controls the last
$(N_P+1)$ spins of this extended chain. Recall that the SVD method
depends on a choice of total communication time $T$. As in the
example earlier, we may wish to search over a range of values of
$T$ to find the most suitable value. It is important for the
procedure at hand that we restrict ourselves to states in the
$\cH^{(1)}$ subspace (whereas before this restriction was just a
way of making the solution much faster to compute). So, we
calculate $\vec{w}_1(T)$, the first right-singular-vector of
$P_{\cB \cap \cH^{(1)}} e^{-i \tilde{H}^{(1)} T} P_{\cA \cap
\cH^{(1)}}$, where $\tilde{H}^{(1)}$ is the part of $\tilde{H}$
that acts on the $\cH^{(1)}$ subspace.

Then, we need to be able to calculate the evolution of the state
$\vec{w}_1$, over a range of times $t$ from $0$ to $T$. Let
$|\psi(0)\rangle$=$\vec{w}_1(T)$, and
$|\psi(t)\rangle=e^{-iH^{(1)}t}|\psi(0)\rangle$. As earlier, the
evolving state is a series of complex coefficients $\psi_j(t)$,
where $j$ is the index to a spin site, ranging from $(1-N_P)$ to
$(N+N_P-1)$. We need to know $\psi_1(t)$ and $\psi_N(t)$ for every
value of $t$ that we wish to calculate $J_A(t)$ and $J_B(t)$ for.

 Similarly we use $\phi_j(t)$, $j=1,\dots,N$ to denote the
evolution of the $|1\rangle$ qubit state over the original
physical chain. Recall that in this scheme, Alice places the qubit
state to be sent, unencoded, onto spin number 1, after all other
spins have been initialised to zero. So, initially we have
$\phi_j(0)=\delta_{j,1}$. The functions $\phi_j(t)$ depend on the
control functions $J_A(t)$ and $J_B(t)$ (whereas the $\psi_j(t)$
do not).

The aim is to chose control functions $J_A(t)$ and $J_B(t)$ in
such a way as to force $\phi_j(t)=\psi_j(t)$, for all the spins in
the range $j=2,\dots,N-1$, and for all $t$ in the interval
$[0,T]$. That is, we know the way the optimal encoded state
evolves over the modified chain, and we want to make the
$|1\rangle$ state in the physical system evolve in exactly the
same way, over all spins except $1$ and $N$. In this way, the
physical system will carry a qubit across it's length with the
same fidelity as the encoded modified system does.

The interactions on the spins from site 3 to site $N-2$ are the
same in the physical chain as in the modified chain. So, the
differential equations for the $\psi_j(t)$ are the same as those
for the $\phi_j(t)$, for $j=3,\dots,N-2$. Specifically,
\begin{eqnarray}
\frac{d\psi_j(t)}{dt} &=& -2i \left[ J_{j-1}\psi_{j-1}(t) +
J_{j}\psi_{j+1}(t) \right] \hspace{0.5cm} \mathrm{and}
\nonumber \\
\frac{d\phi_j(t)}{dt} &=& -2i \left[ J_{j-1}\phi_{j-1}(t) +
J_{j}\phi_{j+1}(t)\right],
\end{eqnarray}
for $j=3,\dots,N-2$. Also, the initial conditions are the same
between the $\psi_j$ and the $\phi_j$, for $j=2,\dots,(N-1)$:
$\psi_j(0)=\phi_j(0)=0$ .

It follows that if we can use our control functions to force
$\frac{d\phi_2(t)}{dt}=\frac{d\psi_2(t)}{dt}$, and
$\frac{d\psi_{N-1}(t)}{dt}=\frac{d\phi_{N-1}(t)}{dt}$, over the
time range $t=0,\dots,T$, then we will have $\psi_j(t)=\phi_j(t)$
for all $t$ in that time range, and for {\em all} $j=2,\dots,N-1$,
as desired.

Now,
\begin{equation}
\frac{d\psi_2(t)}{dt}=-2i\left[ \psi_1(t) + J_2\psi_3(t) \right]
\end{equation}
and
\begin{equation}
\frac{d\phi_2(t)}{dt}=-2i\left[ J_A(t) \phi_1(t) + J_2\phi_3(t)
\right].
\end{equation}

So, assuming that at time $t$ $\phi_j(t)=\psi_j(t)$ for
$j=2,\dots, N-1$, then
$\frac{d\phi_2(t)}{dt}=\frac{d\psi_2(t)}{dt}$ by setting
\begin{equation}
J_A(t)=\frac{\psi_1(t)}{\phi_1(t)}, \eqn{JA}
\end{equation}
and $\frac{d\phi_{N-1}(t)}{dt}=\frac{d\psi_{N-1}(t)}{dt}$ by
setting
\begin{equation}
J_B(t)=\frac{\psi_N(t)}{\phi_N(t)}. \eqn{JB}
\end{equation}

Thus, the practical task of numerically calculating $J_A(t)$ and
$J_B(t)$ involves simulating the evolution of both the $\phi_j$
and $\psi_j$ states on the original and modified systems
respectively, over the time interval $[0,T]$, and evaluating
\eqs{JA} and (\ref{JB}).

%So, if \eq{JA} and \eq{JB} are used for the control functions,
%then the system will communicate with the same fidelity as the
%optimal encoding on the extended system.

The functions $J_A(t)$ and $J_B(t)$ must of course be real-valued,
for the Hamiltonian to be Hermitian. Equations (\ref{JA}) and
(\ref{JB}) will indeed be real for the $XY$ chain. The expressions
for the $\frac{d\phi_j}{dt}$ are all given by a purely imaginary
linear combination of the nearest-neighbour values $\phi_{j-1}$
and $\phi_{j+1}$. Then, considering the initial conditions,
$\phi_j(0)=\delta_{1,j}$, it's clear that the $\phi_j(t)$ are real
for odd $j$ and imaginary for even $j$, for all values of $t$. The
values $\psi_j(t)$ also have this property of alternating real and
imaginary values. Again, the time derivatives of $\psi_j(t)$ are
purely imaginary linear combinations of the values $\psi_{j-1}(t)$
and $\psi_{j+1}(t)$. Thus, by performing the change of variables
\begin{equation}
\psi_j'(t) = \left\{
\begin{array}{ll}
\psi_j(t) & \textrm{if $j$ is
odd, and} \\
i\psi_j(t) \hspace{0.5cm} & \textrm{if $j$ is even,}
\end{array}
\right.
\end{equation}
the differential equations for $\psi'_j(t)$ will all have real
coefficients. So, the entries of the evolution matrix $e^{-i
\tilde{H}^{(1)}T}$ must be real, after that change of variables.
Thus, so must be the entries of $P_{\cB \cap \cH^{(1)}} e^{-i
\tilde{H}^{(1)} T} P_{\cA \cap \cH^{(1)}}$. So, $\vec{w}(T)$,
which is the right-singular vector of $P_{\cB \cap \cH^{(1)}}
e^{-i \tilde{H}^{(1)} T} P_{\cA \cap \cH^{(1)}}$, will also have
all real coefficients with respect to the changed variables.
Changing variables back, the initial encoded state
$\psi_j(0)=\vec{w}_j(T)$ will thus have the property of having
real values for odd $j$ and imaginary values for even $j$, and so
will $\psi_j(t)$ for all $t$. So, \eqs{JA} and (\ref{JB}) will be
real-valued as required.

\eqs{JA} and (\ref{JB}) will never be infinite. In fact,
$|J_A(t)|$ and $|J_B(t)|$ will be at most 1. This is a simple
consequence of conservation of probability. Since
$\psi_j(t)=\phi_j(t)$ over the bulk of the chain ($j=2,\dots,N-1$)
and Alice and Bob's sides only interact via the bulk of the chain
for both the physical and modified systems, we have that
\begin{eqnarray}
\sum_{j=1-N_P}^1 |\psi_j|^2 &=& |\phi_1|^2\textrm{, and} \\
\sum_{j=N}^{N+N_P} |\psi_j|^2 &=& |\phi_N|^2,
\end{eqnarray}
from which it follows that $|\psi_1(t)|\le|\phi_1(t)|$ and
$|\psi_N(t)|\le|\phi_N(t)|$. So, $|J_A(t)|\le 1$ and $|J_B(t)|\le
1$, if they are defined. If $J_A(t)$ (or $J_B(t)$) is undefined
($0/0$), it means that the requirement of
$\frac{d\phi_2(t)}{dt}=\frac{d\psi_2(t)}{dt}$ ( respectively
$\frac{d\phi_{N-1}(t)}{dt}=\frac{d\psi_{N-1}(t)}{dt}$  ) is
satisfied {\em regardless} of the value of $J_A(t)$ (respectively
$J_B(t)$) for that $t$, in which case the value of the control
function can be chosen arbitrarily at that time.

We now plot the derived control functions $J_A(t)$ and $J_B(t)$
for two simple example $XY$ chain systems. We used numerical
integration in these examples, in calculating the evolution of the
$\phi_j(t)$ due to the time-varying Hamiltonian. We divided the
total evolution into a number of discrete time steps, where the
approximation was made that the Hamiltonian remains constant
throughout each step. The value of $J_A(t)$ and $J_B(t)$ for a
step was calculated from the state of the system at the previous
step. We used 2000 time steps, which gave a final fidelity in the
physical chain within two significant figures of the correct value
given by the evolution of the static modified system.

\begin{figure}%[!tbp]
\begin{center}
\epsfxsize=8cm \epsfbox{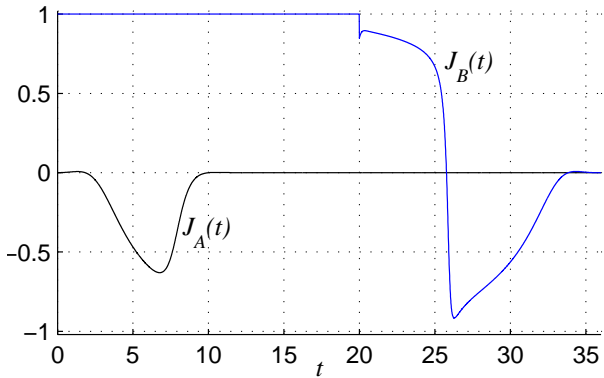}
 \caption{
Control functions for a 104-spin $XY$ chain, where the
non-controlled coupling strengths all have the same strength, 1.
 }
 \label{fig:const}
\end{center}
\end{figure}

 The first example is a
chain 104 spins long (ie.~100 non-controlled spins, plus the four
control spins), with all the non-controlled coupling strengths set
to the same value, 1. The control functions were derived by using
a modified chain 144 spins long (that is, 20 phantom spins added
to each side) with all coupling strengths set to 1, and the total
communication time $T$ chosen to be $36$. Figure (\ref{fig:const})
shows that the resulting $J_A(t)$ and $J_B(t)$ are quite simple
and well behaved. The fidelity measure $\C_B(T)$ is $1.0$, to 6
decimal places. This can be compared with the fidelity in the same
104-spin system but without the time-dependent control, that is
with $J_A(t)$ and $J_B(t)$ fixed at 1: over the time interval
$0<t<1000$, the value of $\C_B(t)$ is at most $0.2809$.

\begin{figure}%[!tbhp]
\begin{center}
\epsfxsize=8cm \epsfbox{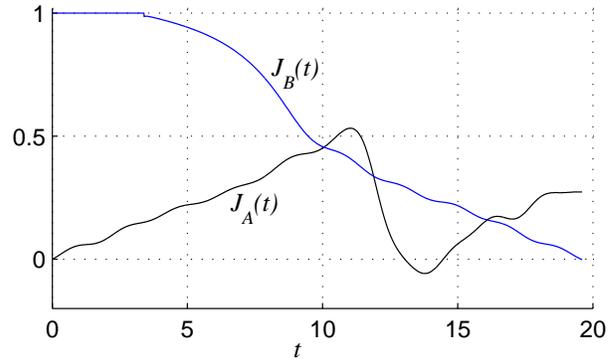}
 \caption{Control functions for a 29-spin $XY$ chain, where the
 non-controlled coupling strengths were chosen randomly from the interval $[0.95,1.05]$.}
 \label{fig:rand}
\end{center}
\end{figure}

The second example is an $XY$ chain 29 spins long, but where the
non-controlled coupling strengths are randomly sampled uniformly
from the interval $[0.95,1.05]$. This is as if the chain has been
manufactured with random imperfections in the coupling strengths,
but these coupling strengths have been somehow measured after the
manufacturing process and are known to Alice and Bob. (A shorter
chain was chosen in this example, compared with the previous
example, in order that a near-perfect fidelity would still result.
We have observed that when random couplings are used, the
achievable fidelity will decrease as a function of the chain
length). We derived control functions using a modified system with
25 phantom spins added to each side, where the new arbitrary
coupling strengths are set to 1. The communication time $T$ was
chosen to be $19.5$. Figure (\ref{fig:rand}) shows control
functions which are a little more complicated in this case, but
still rather smooth. The fidelity measure is $\C_B(T)=0.99625$. In
comparison, in the non-controlled version of this system, with
$J_A(t)=J_B(t)=1$, the value of $\C_B(t)$ does not exceed $0.496$
over the interval $0<t<1000$. Animations of both examples are
available online \cite{Haselgrove04z}.

%%%%%%%%%% BEYOND THE XY CHAIN

What about a system that is not simply an $XY$ chain, but any
configuration conforming to Figure (\ref{fig:d}) and conserving
$\ztot$? Then, the ideas and methods are almost the same, but with
the added complication that we need control the $z$ magnetic
fields on Alice and Bob's qubits as well as controlling $J_A(t)$
and $J_B(t)$. The are two reasons why the magnetic control is
needed, which we explain for Alice's side. First, the simple phase
relation between $\psi_1(t)$ and $\phi_1(t)$ that we saw in the
$XY$ chain does not occur in general. So, $B_{A1}(t)$ is chosen
simply to keep $\phi_1(t)$ in constant relative phase to the
$\psi_1(t)$. Second, the type of interaction that the $J_A(t)$ is
modulating may contain it's own magnetic-field-like interactions
(that is, non-equal diagonal elements in the Hamiltonian) that
need to be cancelled by $B_{A1}(t)$ and $B_{A2}(t)$. General
expressions for $B_{A1}(t)$ and $B_{A2}(t)$ are straightforward to
derive, but not particularly illuminating, so will not be given
here.

%%%%%%%%%%% CONCLUSION
\section{Conclusion} \label{conclusion}

We have considered the problem of communicating a quantum state
over an arbitrary $\ztot$-conserving spin system. Our first scheme
used a static system Hamiltonian, and utilised the fact that the
sender and receiver control several spins each, to increase
fidelity by performing state encoding. We showed that choosing the
optimal state encoding is a simple matter of performing a SVD on a
modified evolution matrix.

 We have also
shown that if the sender and receiver have control of just two
spins each, but can vary the interactions on these four spins over
time, then they can achieve a fidelity that is equal to if they
each controlled many more spins on a static system and used the
optimal state encoding. We have given a practical method of
deriving suitable control functions. The advantage of this scheme
is the ``fixed interface'' that Alice and Bob have with the chain.
That is, if the chain is altered, the only change that Alice and
Bob need make is to their control functions, rather than to the
number of spins they control.

It should be noted that the systems we have considered are
idealised to a high degree. In particular, we haven't considered
the effects of external noise, or the effect of having a
Hamiltonian that only {\em approximately} commutes with $\ztot$,
or the case where Alice and Bob have only an approximate knowledge
of the system Hamiltonian. These issues will be the subject of
future work by the author.

\section*{Acknowledgements}
I would like to thank Tobias Osborne and Michael Nielsen for their
detailed comments on the original manuscript, and for helpful and
enlightening discussions relating to this work.

\appendix
\section{}
Here we outline a proof of the claim in Sec.~\ref{firstsec}
regarding the connection between $\C_B(T)$ and the system's
ability to transmit entanglement from Alice to Bob. This
connection helps establish $\C_B(T)$ as a good measure of
communication fidelity.

\begin{proposition}
Suppose that Alice sends a state which is maximally entangled with
some additional spin that Alice possesses. (After the maximally
entangled state is created, the additional spin is assumed to not
interact during the remainder of the communication procedure).
Then, after the communication procedure of Sec.~\ref{firstsec} is
carried out, the entanglement (measured by concurrence) between
Alice's additional spin and Bob's decoded message, equals
$\sqrt{\C_B(T)}$.
\end{proposition}

\textbf{Proof:} Note that it doesn't matter which
maximally-entangled state is used --- all such states are
equivalent up to a local unitary on the additional spin, and such
a local unitary could not possibly affect the way entanglement is
transferred through the system.

Let the additional spin ``$\ad$'' and the spin ``$M$'' containing
the message have the maximally entangled state
$\frac{1}{\sqrt{2}}(|0\rangle_{\ad}|0\rangle_M+|1\rangle_{\ad}|1\rangle_M$).
Thus, after Alice performs her encoding, the entire state is:
\begin{equation}
|\Phi(0)\rangle=\frac{1}{\sqrt{2}}\left[
|0\rangle_{\ad}|\mathbf{0}\rangle_A|\mathbf{0}\rangle_{\bar{A}} +
|1\rangle_{\ad}|1_{ENC}\rangle_A|\mathbf{0}\rangle_{\bar{A}}\right].
\end{equation}
After the system evolves for time $T$, the state becomes
\begin{eqnarray}
|\Phi(T)\rangle&=&\frac{1}{\sqrt{2}}\Big[
|0\rangle_{\ad}|\mathbf{0}\rangle_{\bar{B}}|\mathbf{0}\rangle_{B}
+ |1\rangle_{\ad} ( \sqrt{1-\C_B(T)}|\eta(T)\rangle \nonumber\\
&&+ \sqrt{\C_B(T)}|\mathbf{0}\rangle_{\bar{B}}|\gamma(T)
\rangle_B) \Big].
\end{eqnarray}
Then Bob performs a decoding unitary, denoted $\udec$, on the
spins he controls. $\udec$ is defined to act as follows: $\udec
|\mathbf{0}\rangle_B = |\mathbf{0}\rangle_B$ and $\udec
|\gamma(T)\rangle_B = |0\dots01\rangle_B$, where
$|0\dots01\rangle_B$ is the $|1\rangle$ state on spin $N$ and the
all-zero state on Bob's other spins. After Bob's decoding, the
joint state of Alice's additional spin and Bob's decoded spin is:
\begin{eqnarray}
\rho_{\ad/N}&=&\tr_{\overline{\ad/N}}(\udec
\tr_{\bar{B}}(|\Phi(T)\rangle\langle\Phi(T)|)\udec^\dagger)
\nonumber \\
&=& \frac{1}{2}\Big[(1-\C_B(T))|1\rangle\langle 1|\otimes
\rhotilde
 + \C_B(T)|11\rangle\langle 11|
\nonumber\\
&&\hspace{1em}+ |00\rangle\langle00|+ \sqrt{\C_B(T)} (
|00\rangle\langle11|
\nonumber\\&&
 \hspace{1em}+ |11\rangle\langle00|)\Big],
\end{eqnarray}
where $\rhotilde\equiv\tr_{\overline{\ad/N}}(\udec
\tr_{\bar{B}}(|\eta(T)\rangle\langle\eta(T)|)\udec^\dagger)$, and
where $\tr_{(\cdot)}(\cdot)$ is the partial trace performed over
the spins indicated.

Concurrence is a measure of entanglement between two qubits
\cite{Wootters98a}. The value of concurrence for a density matrix
$\rho_{\ad/N}$ is equal to
\begin{equation}
E(\rho_{\ad/N}) = \max\{ 0,
\sqrt{\lambda_1}-\sqrt{\lambda_2}-\sqrt{\lambda_3}-\sqrt{\lambda_4}\},
\end{equation}
where the $\lambda_j$s are the eigenvalues, in nonincreasing
order, of the matrix
$\rho_{\ad/N}(\sigma^y\otimes\sigma^y)\rho_{\ad/N}^*(\sigma^y\otimes\sigma^y)$,
where $^*$ represents complex conjugation in the computational
basis. It can be shown that
\begin{eqnarray}
\lambda_1&=&\frac{1}{4}\left(\sqrt{\rhotilde_{11}\C_B(T) + 1 - \rhotilde_{11}}+\sqrt{\C_B(T)}\right)^2 \nonumber\\
\lambda_2&=&\frac{1}{4}\left(\sqrt{\rhotilde_{11}\C_B(T) + 1 -
\rhotilde_{11}}-\sqrt{\C_B(T)}\right)^2
 \nonumber\\
\lambda_3&=&0 \nonumber\\
\lambda_4&=&0,
\end{eqnarray}
where $\rhotilde_{11}=\langle 0 | \rhotilde | 0\rangle$. Thus,
using the fact that $\C_B(T)$ and $\rhotilde_{11}$ each lie in the
interval $[0,1]$, we have
\begin{equation}
E(\rho_{\ad/N}) = \sqrt{\C_B(T)},
\end{equation}
as required.
 \qed

%\bibliography{mybib}

\end{document}